\documentclass{Interspeech2024}
\usepackage{booktabs}
\usepackage{subfigure}
\usepackage{multirow}
\usepackage{array}
\usepackage{graphicx}
\usepackage{hyperref}

\interspeechcameraready

\title{SE/BN Adapter: Parametric Efficient Domain Adaptation for Speaker Recognition}

\name[affiliation={1,2}]{Tianhao}{Wang}
\name[affiliation={1}]{Lantian}{Li}
\name[affiliation={2}]{Dong}{Wang}

\address{
  $^1$School of Artificial Intelligence, Beijing University of Posts and Telecommunications, China \\
  $^2$Center for Speech and Language Technologies, BNRist, Tsinghua University, China
  \thanks{This work was supported by the National Natural Science Foundation of China (NSFC) under Grants No.62301075/62171250.}}
\email{Corresponding authors:~lilt@bupt.edu.cn, wangdong99@mails.tsinghua.edu.cn}


\begin{document}

\maketitle

\begin{abstract}

Deploying a well-optimized pre-trained speaker recognition model in a new domain often leads to a significant decline in performance. While fine-tuning is a commonly employed solution, it demands ample adaptation data and suffers from parameter inefficiency, rendering it impractical for real-world applications with limited data available for model adaptation. Drawing inspiration from the success of adapters in self-supervised pre-trained models, this paper introduces a SE/BN adapter to address this challenge. By freezing the core speaker encoder and adjusting the feature maps' weights and activation distributions, we introduce a novel adapter utilizing trainable squeeze-and-excitation (SE) blocks and batch normalization (BN) layers, termed SE/BN adapter. Our experiments, conducted using VoxCeleb for pre-training and 4 genres from CN-Celeb for adaptation, demonstrate that the SE/BN adapter offers significant performance improvement over the baseline and competes with the vanilla fine-tuning approach by tuning just 1\% of the parameters.

\end{abstract}

\section{Introduction}

The real-world deployment of speaker recognition models often suffers from significant performance degradation due to the variability of acoustic environments and limited training data coverage. This mismatch between training and deployment domains is a well-known challenge~\cite{huh2023voxsrc,sadjadi20222021}. 
Existing adaptation techniques aim to adapt a pre-trained model to a new domain. These approaches can be categorized into (1) Front-end embedding network adaptation, fine-tuning the speaker embedding network to align with the target domain's embedding distribution~\cite{zhang2018analysis,
wang2018unsupervised,tu2019variational,chen2020adversarial,huang2024adapter}.
(2) Back-end scoring model adaptation, modifying the parameters of the scoring model while keeping the 
front-end network unchanged~\cite{garcia2014supervised,garcia2014unsupervised,sun2016return,sun2016deep,
lee2019coral+,wang2020generalized}.

Recent focus has shifted towards front-end adaptation due to the success of simple cosine back-ends with modern speaker recognition models trained using margin-based loss functions~\cite{xiang2019margin, wang2022scoring}. This eliminates the need for complex back-end models like PLDA~\cite{ioffe2006probabilistic}.
In practice, front-end adaptation can be achieved through fine-tuning the entire model with domain-specific data. However, real-world scenarios often have limited domain-specific data, e.g., with only 3-5 recordings per person for tens to hundreds of individuals. Fine-tuning the entire model with such scarce data often leads to severe overfitting. Moreover, it also lacks parameter efficiency, as a new set of parameters is required for each domain.

The adapter concept, initially introduced in computer vision and NLP~\cite{rebuffi2017learning,pfeiffer2020adapterhub,he2021towards}, tackles these limitations by augmenting the core model with extra concise structures and altering the core model's behaviour by adjusting the augmented structures.
The original aim was to adjust large-scale self-supervised learning (SSL) models for downstream tasks. Due to the large number of parameters in SSL models, updating the entire set is inefficient and risky. Adapters freeze the pre-trained model and insert lightweight, adaptable structures (e.g., linear transforms) between intermediate layers. This structure called an adapter, promotes parameter efficiency and avoids overfitting due to its limited parameters and modular design.
The success of adapters on SSL models in speech recognition tasks like speech recognition and translation~\cite{ otake2023parameter, thomas2022efficient, huang2023findadaptnet, le2021lightweight} using models like wav2vec 2.0~\cite{baevski2020wav2vec}, HuBERT~\cite{hsu2021hubert}, and WavLM~\cite{chen2022wavlm} has been significant.

Inspired by the success of adapters in SSL models, this paper introduces an adapter designed for conducting domain adaptation tasks in speaker recognition, specifically focusing on challenges related to low-resource domain adaptation. Our goal is to adjust a pre-trained speaker model (not SSL) to a new domain with limited labelled data by updating only a small subset of network parameters while keeping the core speaker embedding network fixed. To achieve this objective, we propose a novel ultra-lightweight domain adapter.

The development of such an adapter requires some prior knowledge. Our premise is that the fundamental speaker patterns have been effectively learned by the speaker embedding network through extensive training data, and domain mismatch can be largely attributed to subtle differences when lower-level patterns are combined into higher-level patterns. For example, in certain domains, certain patterns may be emphasized and therefore require more attention; conversely, in other domains, these patterns may need to be diminished or even disregarded. 
If this premise holds, the adaptation can be easily implemented by adjusting the weights of the channel maps and normalizing the distributions of the feature maps. We actualize this concept by utilizing the squeeze-and-excitation (SE) module, where the linear projections are adjustable. Additionally, Sarfjoo et al.~\cite{sarfjoo2020supervised} have discovered that batch-norm (BN) is also a lightweight structure that can be utilized for low-resource adaptation by tuning the mean shift and covariance. Through our experiments, we found that SE and BN adaptation complement each other well, and their combination results in strong performance. Therefore, we have devised a lightweight SE/BN adapter that incorporates both SE and BN, with parameters constituting only 1\% of the full network but capable of delivering comparable or even superior performance compared to full-model fine-tuning.



\section{Related Work}

Research on front-end adaptation has given rise to numerous methods, broadly categorized into unsupervised and supervised modes. Unsupervised adaptation methods do not require labelled speaker identities for the target domain speech and primarily focus on aligning the marginal distributions between the source and target domains. Two common approaches are distribution alignment and domain adversarial learning. Distribution alignment aims to minimize the differences between domains to acquire domain-invariant representations~\cite{lin2020multi,hu2022class,zhou2023investigation}, while domain adversarial learning employs a gradient reversal layer to decrease domain discrepancies and unify diverse domain data within a shared subspace~\cite{wang2018unsupervised,ganin2015unsupervised,wang2021adversarial}.

In contrast, supervised adaptation methods necessitate speaker labels. The straightforward approach involves fine-tuning the entire pre-trained model~\cite{dubey2019transfer}; however, more effective strategies leverage prior knowledge of the network, in-domain data, and characteristics of the domain shift. For example, Sarfjoo et al.~\cite{sarfjoo2020supervised} discovered that shallow layers of TDNN models are closely linked to domain specificity, and this specificity potentially manifests in the statistical properties (shift and variance) of the channel activities. However, our experiments show that BN adapters cannot beat a strong fine-tuned model.

Another relevant study comes from Huang et al.~\cite{huang2024adapter}, where the authors similarly developed a lightweight adapter to address domain discrepancies. Their adapter's primary objective is to incorporate a domain descriptor (such as a one-hot indicator or embedding) to modify the model's behaviour, leading to the ability to handle unseen domains. In contrast to their work, we aim to adapt a pre-trained model to a specific domain with limited adaptation data and minimal parameter adjustments. 

\section{SE/BN Adapter}

\subsection{Revisit SE}

\begin{figure}[ht]
\centering
\includegraphics[width=\linewidth]{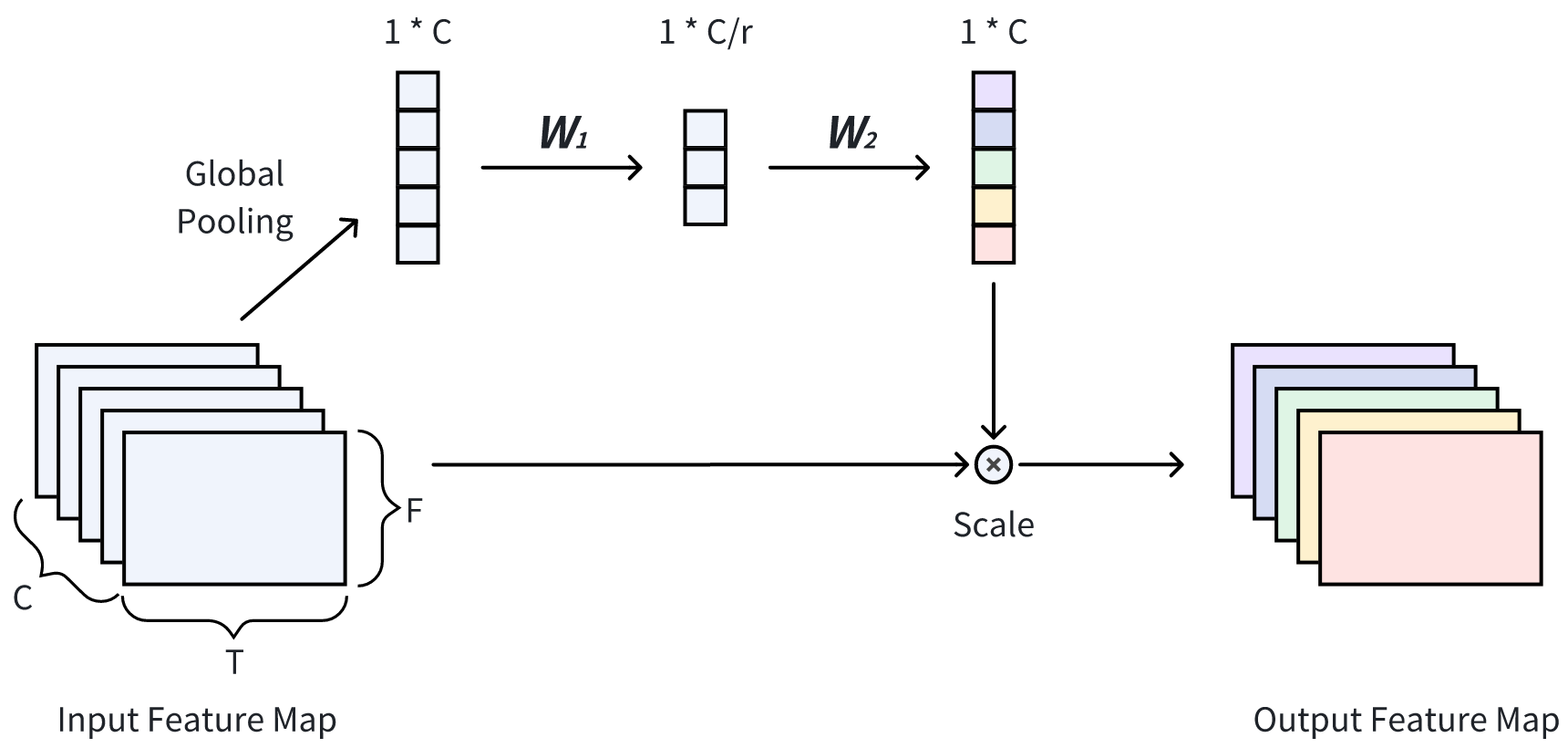}
\caption{Illustration of the Squeeze-and-Excitation (SE) block. The excitation function $f=\sigma (\mathbf{W}_2 \delta (\mathbf{W}_1 \mathbf{z}))$ is depicted, where $\mathbf{W}_1$ and $\mathbf{W}_2$ matrices are shown.}
\label{fig:se_block}
\end{figure}
\vspace{-1mm}

The Squeeze-and-Excitation (SE) block was initially introduced to enhance CNN-based image classification~\cite{hu2018squeeze} and later successfully applied to speaker recognition~\cite{desplanques2020ecapa,rouvier2021studying}. The fundamental concept behind this innovation is to capture global information from all channels' feature maps and utilize this information to amplify features generated by local convolutions.

An SE block consists of two primary components: Squeeze and Excitation. We will explain these components in the context of speaker recognition, as shown in Figure~\ref{fig:se_block}. The Squeeze component employs a Global Average Pooling (GAP) operation to compute a global activation value for each feature map, thereby establishing a global receptive field for each feature map. Formally, considering input feature maps denoted as $\mathbf{X}\in \mathbb{R}^{C\times F\times T}$, where $C$, $F$, and $T$ represent the number of channels, frequency bins, and frames, respectively. Each channel corresponds to a feature map, represented as a $F \times T$ matrix. The output $\mathbf{z}_c$ from GAP for the $c$-th channel can be calculated as:

\begin{equation}
\mathbf{z}_c=\frac{1}{F\times T}\sum^F_{i=1}\sum^T_{j=1}\mathbf{X}_c(i,j)
\end{equation}

The Excitation component transforms $\mathbf{z}$ into a scaling vector $\mathbf{s}$ through two linear transformations:

\begin{equation}
\mathbf{s} = \sigma (\mathbf{W}_2 \delta (\mathbf{W}_1 \mathbf{z}))
\end{equation}

\noindent where $\delta$ represents the ReLU function, $\mathbf{W}_1 \in \mathbb{R}^{C/r \times C}$ and $\mathbf{W}_2 \in \mathbb{R}^{C \times C/r}$, and $\sigma$ denotes the sigmoid function. Finally, the original feature maps are scaled by $\mathbf{s}$ to generate the final feature maps, expressed as:

\begin{equation}
\tilde{\mathbf{X}} = \mathbf{X} \cdot \mathbf{s}
\end{equation}

\subsection{SE layer is domain specific}

To further understand the behaviour of SE blocks, we trained two deep speaker models, ResNet34 and ResNet34SE, using VoxCeleb2.dev as the training dataset. Notably, both models employed the same input features, pooling strategy, and loss functions. The key distinction lies in ResNet34SE integrating an SE block after each ResNet block that involves two CNN layers. These models were evaluated on two in-domain test sets, Vox1-E and Vox1-H, as well as four out-of-domain test sets: CNC.e, CNC.i, CNC.l, and CNC.s. The out-of-domain test sets are subsets of the CNCeleb evaluation set, representing domains related to entertainment, interview, live\_broadcast, and singing, respectively. Additional details regarding these datasets can be found in Section 4. The results in terms of equal error rate (EER) are summarized in Table~\ref{tab:pretrained}.

\begin{table}[htb!]
\centering
\caption{Comparison of EER (\%) between ResNet34 and ResNet34SE.}
\label{tab:pretrained}
\resizebox{1.0\columnwidth}{!}{
\begin{tabular}{lcccccc}
\toprule
Model      & Vox1-E         & Vox1-H         & CNC.e            & CNC.i           & CNC.l          & CNC.s  \\
\cmidrule(r){1-1} \cmidrule(r){2-3} \cmidrule(r){4-7}
ResNet34   & 1.244          & 2.217          & \textbf{11.017}  & \textbf{8.365}   & \textbf{6.163} & \textbf{25.924} \\
ResNet34SE & \textbf{1.214} & \textbf{2.215} & 11.383           & 8.396            & 6.467          & 28.388 \\
\bottomrule
\end{tabular}}
\end{table}

The results indicate that ResNet34SE consistently outperforms ResNet34 in the two in-domain tests but falls short in all four out-of-domain tests. This suggests that integrating the SE block enhances the model's performance on the training data but compromises its ability to generalize across different domains. This observation supports the hypothesis that SE blocks are domain-specific and should be carefully adjusted when adapting a pre-trained model to new domains. With this understanding, one can devise an efficient and lightweight adaptation strategy by updating the parameters of the SE block while maintaining the core structure of the backbone network unchanged. This concept inspired our design of a domain adapter based on SE blocks.

\subsection{SE/BN Adapter}

\begin{figure}[ht]
\centering
\includegraphics[width=\linewidth]{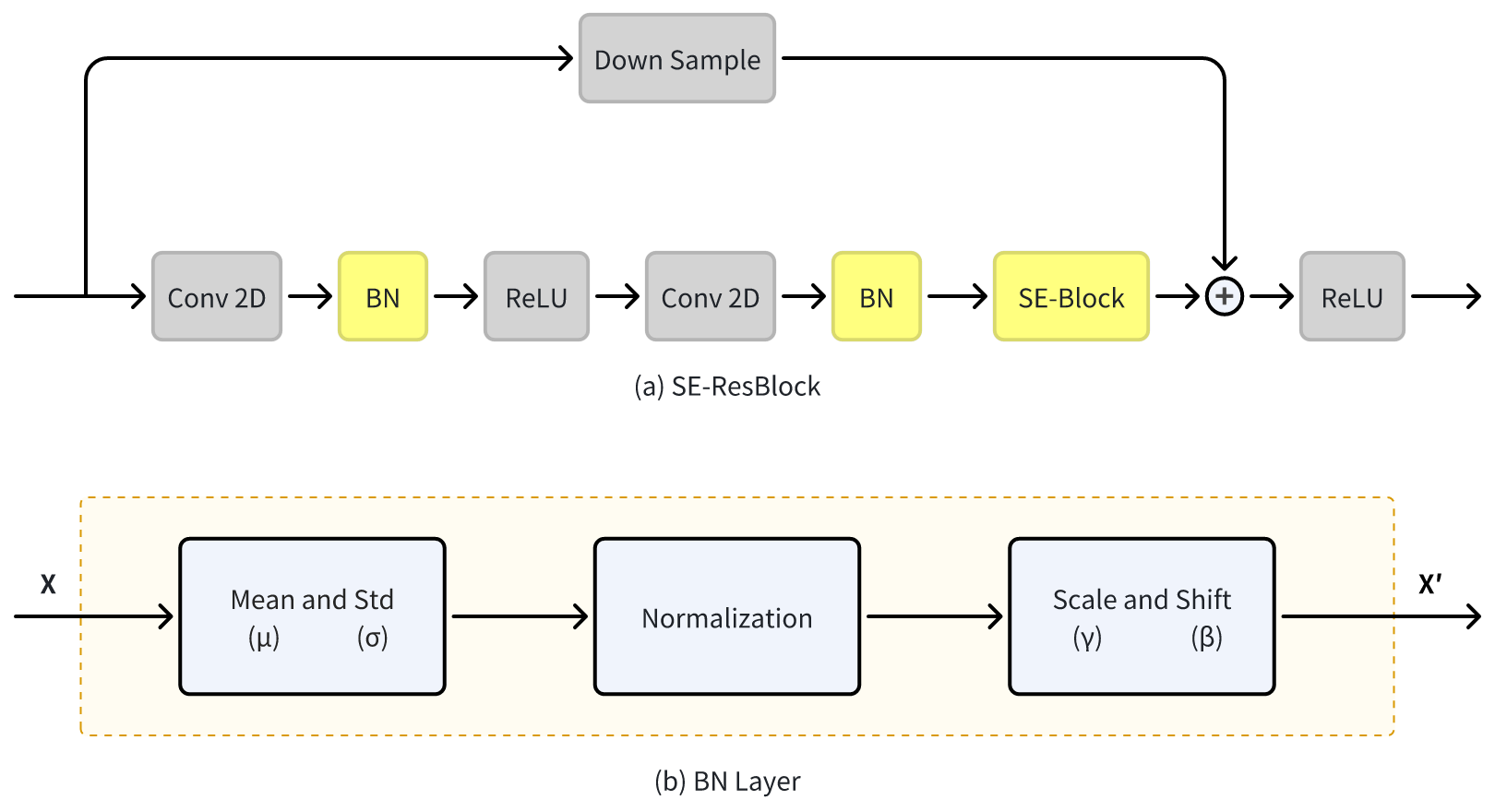}
\caption{SE/BN adapter based on SE and BN. (a) A ResNet block with SE blocks and BN layers. (b) BN layer. The yellow color indicates trainable parameters.}
\label{fig:adapter}
\end{figure}

Taking ResNet as an example, we integrate the SE block after each ResNet block to rescale the output feature maps and enable the adaptation of SE block parameters using data from the new domain. This adaptation is referred to as the \emph{SE adapter}. Additionally, we explored a domain adaptation approach based on BN layers, named \emph{BN adapter}~\cite{sarfjoo2020supervised}. 
Batch normalization, as illustrated in Figure~\ref{fig:adapter}(b), can be simply formulated as follows:

\begin{equation}
\mathbf{X}' = \beta + \gamma \frac{\mathbf{X} -\mu } {\sigma}
\end{equation}
\noindent where $\mu$ and $\sigma$ represent the mean and standard deviation of the current batch, and $\gamma$ and $\beta$ are two adjustable parameters. Tuning these parameters can compensate for the global shift and scaling caused by domain mismatch. 

Since the SE adapter and BN adapter tune models in different ways, they can be combined to construct a more powerful adapter, denoted by \emph{SE/BN adapter}. The framework is illustrated in Figure~\ref{fig:adapter}(a), where the grey blocks are frozen, and the yellow blocks are trainable. Notably, compared to the primary ResNet blocks, the parameters of SE blocks and BN layers are negligible, offering an extremely lightweight domain adaptation approach.

\section{Experiments}

\subsection{Data}

Our experiments utilized two datasets: VoxCeleb~\cite{nagrani2020voxceleb} and CN-Celeb~\cite{li2022cn}. Specifically, the development set of the VoxCeleb2 dataset, comprising a total of 5,994 speakers, was employed to establish the pre-trained model. For domain adaptation, we selected four subsets from the CN-Celeb1.dev and CN-Celeb2 datasets, corresponding to four genres (entertainment, interview, live\_broadcast, singing) with the highest speaker counts. These genres involve different characteristics in the acoustic environment and speaker style, therefore representing different domains. Performance evaluation was carried out using four subsets of the CN-Celeb1.eval dataset, aligning with the four genres in the adaptation data. The data breakdown for these four genres is detailed in Table~\ref{tab:data}.

\begin{table}[htb!]
\centering
\caption{Data Profile of Four Genres in CN-Celeb.}
\label{tab:data}
\resizebox{1.0\columnwidth}{!}{
\begin{tabular}{lccccc}
\toprule
\multirow{2}{*}{\textbf{Genre}} & \multicolumn{2}{c}{\textbf{Dev Set}} & \multicolumn{3}{c}{\textbf{Test Set}} \\
& Spks   & Utters      & Spks   & Utters   & Trials   \\
\cmidrule(r){1-1} \cmidrule(r){2-3} \cmidrule(r){4-6}
Entertainment   & 975    & 50,780      & 136    & 3,694    & 473,613    \\
Interview       & 1,167  & 88,307      & 149    & 6,521    & 930,750    \\
Live\_broadcast & 480    & 173,525     & 43     & 2,347    & 87,742     \\
Singing         & 645    & 52,709      & 69     & 2,017    & 128,766    \\
\bottomrule
\end{tabular}}
\end{table}

\subsection{Settings}

We followed the voxceleb/v2 recipe in the Sunine toolkit\footnote{https://gitlab.com/csltstu/sunine} to build the pre-trained model. The backbone structure used was ResNet34SE, with an SE block added after each ResNet block. An attentive statistics pooling was employed to generate utterance-level representations, which were then transformed by a fully connected layer to produce 256-dimensional x-vectors. The model was trained using AAM-Softmax with a margin value of 0.2 and a scale factor of 32. Various advanced training techniques were applied, including data augmentation, margin scheduler, and Adam optimizer with learning rate warm-up. Further details can be found in the Sunine repository, with the efficacy of these techniques discussed in~\cite{chen23m_interspeech}. The simple cosine distance was used to score the trials.

\subsection{Basic Results with Adapters}

We initially examined the performance of the SE adapter at different locations within the ResNet34SE architecture to identify which parts of the network are sensitive to domain changes. 
ResNet34SE comprises 16 groups, each containing 3, 4, 6, and 3 ResNetSE blocks from shallow to deep layers. We evaluated performance by adapting the SE blocks in each group individually or across all groups. Results with BN and SE/BN adapters employed in all the groups were also examined. The results in terms of EER are presented in Table~\ref{tab:block}. It should be noted that the number of parameters varies in different tests due to the differing numbers of ResNetSE blocks in each group and the varying numbers of feature channels, with deeper blocks containing more channels. Several key observations are highlighted below.

\begin{table}[htb!]
\centering
\caption{Performance comparison with SE adapters in individual groups and all groups. `Pre-train' and `Fine-tune' denote results with the pre-trained model and a model obtained through full-model fine-tuning. `SE @ Gx' and `SE @ G1-G4' indicate SE adaptation in ResNetSE group x and all groups, respectively. Results with BN adapters and SE/BN adapters are also reported.}
\label{tab:block}
\resizebox{1.0\columnwidth}{!}{
\begin{tabular}{lccccc}
\toprule
EER (\%)         & \# Params     & CNC.e     & CNC.i    & CNC.l     & CNC.s  \\
\cmidrule(r){1-1} \cmidrule(r){2-2} \cmidrule(r){3-6}
Pre-train       &   -           & 11.383    & 8.396     & 6.467     & 28.388 \\
Fine-tune       & 8.0 M         & 7.701     & 5.650     & 5.122     & 17.146 \\
\cmidrule(r){1-1} \cmidrule(r){2-2} \cmidrule(r){3-6}
SE @ G1         & 0.9 K         & 10.540    & 7.046     & 5.556     & 29.107 \\
SE @ G2         & 4.4 K         & 9.837     & 6.999     & 5.512     & 25.719 \\
SE @ G3         & 25.4 K        & 10.118    & 6.937     & 5.686     & 26.437 \\
SE @ G4         & 50.0 K        & 9.753     & 6.764     & 5.859     & 27.413 \\
SE @ G1-G4      & 80.7 K        & 8.994     & 6.701     & 5.556     & 22.947 \\
\cmidrule(r){1-1} \cmidrule(r){2-2} \cmidrule(r){3-6}
BN @ G1-G4              & 7.6 K         & 8.375     & 6.246     & 5.382     & 21.253 \\
SE/BN @ G1-G4         & 88.3 K        & 8.010     & 6.136     & 5.295     & 20.021 \\
\bottomrule
\end{tabular}}
\end{table}

\begin{itemize}

\item The SE adapter consistently outperformed the pre-trained model across all tested domains, irrespective of the location of adaptation within the ResNet34SE architecture, supporting our assumption that domain mismatch can be alleviated by adjusting the pattern strengths at each layer.

\item The optimal group for SE adapter to reside varied across domains. For interview, adaptation in deeper groups was slightly more effective than in shallower groups, whereas for live\_broadcast and singing, adapting middle groups (G2 and G3) yielded better results. This variance could be attributed to complex variations in acoustic conditions and speaking styles across domains, as well as the number of adaptable parameters in each group.

\item Employing SE adaptation across all groups (SE @ G1-G4) outperformed adaptation in individual groups, indicating that performance gains from individual group adaptations are complementary and cumulative.

\item The BN adapter (BN @ G1-G4) consistently outperforms the SE adapter (SE @ G1-G4) with fewer adaptable parameters, showcasing superior performance. Moreover, combining both adapters (SE/BN @ G1-G4) results in further enhancements. This underscores the complexity of domain mismatch, suggesting that it should be addressed through various means such as updated pattern composition (SE adapter) and global shifting and scaling (BN adapter).

\end{itemize}

It can be seen that there is still a performance gap between various adapters and global fine-tuning. However, the difference is marginal compared to the improvement achieved by the adapters over the pre-trained model. The advantage of fine-tuning is attributed to the abundance of adaptation data in this experiment, enabling feasible whole-model fine-tuning. However, SE or BN adaptation achieved comparable results with significantly fewer parameters. For example, the SE/BN adapter involves only 88.3K parameters, representing only 1\% of the entire network (8.0M). In other words, a machine that supports an extra fine-tuning model can support 100 new domains by using the SE/BN adapter.

\subsection{Results with Limited Data}

In this section, we replicate a low-resource scenario commonly encountered in real-world applications. In such conditions, only a limited amount of data from the new domain is available, and fine-tuning the entire model often results in significant overfitting. Initially, we create development sets of varying sizes to simulate resource availability differences and conduct closed-set speaker verification tests. As depicted in Table~\ref{tab:real}, for each genre, four sets of development (enroll) / test data were established, comprising 50, 100, 200, and 400 speakers respectively. Within a development set, up to 5 utterances were randomly selected for each speaker. This set served two purposes: (1) training the three kinds of adapters and (2) constructing speaker enrollment by averaging the embedding vectors of all utterances from the same speaker. The test set was formed by randomly selecting 5-10 utterances for each speaker in the corresponding development set, ensuring no overlap between utterances in the two sets. An enroll-test cross-pairing method was then utilized to generate test trials.

\begin{table}[htb!]
    \centering
    \caption{Development and test data utilized in low-resource testing.}
    \vspace{-1mm}
    \label{tab:real}
    \resizebox{1.0\columnwidth}{!}{
    \begin{tabular}{llcccc}
        \toprule
        \textbf{Genre}                   &  Set  & 50 Spks  & 100 Spks &  200 Spks &  400 Spks \\
        \cmidrule(r){1-1} \cmidrule(r){2-2} \cmidrule(r){3-6}
        \multirow{2}{*}{Entertainment}   & Dev   & 240 & 470 & 944 & 1,855 \\
        & Test  & 260 & 530 & 1,056 & 2,145 \\
        \cmidrule(r){1-1} \cmidrule(r){2-2} \cmidrule(r){3-6}
        \multirow{2}{*}{Interview}       & Dev & 226 & 454 & 936 & 1,816 \\
        & Test & 274 & 546 & 1,064 & 2,184 \\
        \cmidrule(r){1-1} \cmidrule(r){2-2} \cmidrule(r){3-6}
        \multirow{2}{*}{Live\_broadcast} & Dev & 220 & 461 & 901 & 1,836 \\
        & Test & 280 & 539 & 1,099 & 2,164 \\
        \cmidrule(r){1-1} \cmidrule(r){2-2} \cmidrule(r){3-6}
        \multirow{2}{*}{Singing}         & Dev & 227 & 458 & 914 & 1,844 \\
        & Test & 273 & 542 & 1,086 & 2,156 \\
        \bottomrule
    \end{tabular}}
\end{table}

\vspace{-1mm}

Due to the limited adaptation data, utilizing Softmax loss for model adaptation becomes impractical. The primary concern is that the final classification layer would absorb a significant portion of the error signals, leading to gradient vanishing issues for the backbone parameters. Our goal is to minimize loss by adjusting the adaptation parameters rather than the new randomly initialized classification layer. To tackle this challenge, we chose to implement the generalized end-to-end (GE2E) loss function during adaptation~\cite{wan2018generalized}. It is important to highlight that this issue did not manifest in the previous experiment due to the availability of ample adaptation data. The EER results are summarized in Table~\ref{tab:exps}.

\begin{table}[htb!]
    \centering
    \caption{Results in EER (\%) with different adaptation methods in low-resource conditions.}
    \vspace{-1mm}
    \label{tab:exps}
    \resizebox{1.0\columnwidth}{!}{
    \begin{tabular}{clccccc}
        \toprule
        \# Spks                & Method      &  \# Params  & CNC.e & CNC.i & CNC.l & CNC.s \\
        \cmidrule(r){1-1} \cmidrule(r){2-2} \cmidrule(r){3-3} \cmidrule(r){4-7}
        \multirow{5}{*}{50}    & Pre-train  & -         & 4.615 & 5.474 & 6.786 & 16.850    \\
                               & Fine-tune  & 8.0 M     & 3.077 & 3.650 & 5.000 & 12.454    \\
                               & SE         & 80.7 K    & 3.462 & 3.650 & 5.357 & 12.088    \\
                               & BN         & 7.6 K     & 3.462 & 3.650 & 5.000 & 12.088    \\
                               & SE/BN      & 88.3 K    & \textbf{2.692} & \textbf{3.285} 
                                                        & \textbf{4.643} & \textbf{11.722} \\
        \cmidrule(r){1-1} \cmidrule(r){2-2} \cmidrule(r){3-3} \cmidrule(r){4-7}
        \multirow{5}{*}{100}   & Pre-train  & -         & 6.792 & 3.114 & 5.937 & 15.498 \\
                               & Fine-tune  & 8.0 M     & \textbf{6.226} & 2.747 & 3.896 & 12.915 \\
                               & SE         & 80.7 K    & \textbf{6.226} & 2.747 & 3.896 & 13.284 \\
                               & BN         & 7.6 K     & 6.604 & \textbf{2.564} & 4.082 & 13.100 \\
                               & SE/BN      & 88.3 K    & 6.415 & \textbf{2.564} 
                                                        & \textbf{3.525} & \textbf{12.177} \\
        \cmidrule(r){1-1} \cmidrule(r){2-2} \cmidrule(r){3-3} \cmidrule(r){4-7}
        \multirow{5}{*}{200}   & Pre-train  & -         & 5.398 & 5.263 & 5.187 & 15.285 \\
                               & Fine-tune  & 8.0 M     & \textbf{4.072} & 3.759 
                                                        & 3.822 & \textbf{10.958} \\
                               & SE         & 80.7 K    & 4.261 & \textbf{3.665} & 3.731 & 12.523 \\
                               & BN         & 7.6 K     & 4.167 & 3.759 & 3.822 & 11.786 \\
                               & SE/BN      & 88.3 K    & \textbf{4.072} & \textbf{3.665} 
                                                        & \textbf{3.458} & 11.510 \\
        \cmidrule(r){1-1} \cmidrule(r){2-2} \cmidrule(r){3-3} \cmidrule(r){4-7}
        \multirow{5}{*}{400}   & Pre-train  & -         & 5.967 & 4.533 & 5.730 & 15.584 \\
                               & Fine-tune  & 8.0 M     & \textbf{4.802} & \textbf{3.571} 
                                                        & \textbf{3.743} & \textbf{10.158} \\
                               & SE         & 80.7 K    & 4.895 & \textbf{3.571} & 4.251 & 12.291 \\
                               & BN         & 7.6 K     & 4.988 & 3.663 & 3.974 & 11.781 \\
                               & SE/BN      & 88.3 K    & 4.895 & 3.617 & 3.882 & 11.132 \\
        \bottomrule
    \end{tabular}}
\end{table}


The results in Table~\ref{tab:exps} show a clear trend that in scenarios with limited data (e.g., 50 or 100 speakers), the SE/BN adapter tends to outperform the fine-tuning approach with a large margin, though none of the individual adapter beats fine-tuning. This double confirmed that the two types of adapters are complementary. It should be noted that fine-tuning, although shows reasonable performance gain in low-resource conditions, requires tedious learning rate calibration to circumvent issues like loss oscillation and unpredictable outcomes. Conversely, training the adapters exhibits good stability during optimization. When more data is available, e.g., 400 speakers, fine-tuning is a favored approach.

\section{Conclusion}

This paper addresses the challenge of training-deployment mismatch in speaker recognition by introducing an SE/BN adapter, a lightweight and parametric efficient adaptation approach. The underlying hypothesis driving the SE/BN adapter design is that domain mismatch in speaker recognition is not rooted in the speaker patterns themselves, but rather in how these patterns are located, weighted, and integrated. This issue can be mitigated by adjusting the SE blocks and the BN layer. Experimental results demonstrated that the SE/BN adapter outperforms the pre-trained model by a large margin, and even beats the strong fine-tuning model in low-resource conditions with 1\% domain-specific parameters. Future research endeavours will involve testing this approach on diverse datasets and various network structures. Additionally, exploring and elucidating the phenomenon of distributional drift when transitioning between domains will be pursued as another research avenue.

\newpage

\bibliographystyle{IEEEtran}
\bibliography{mybib}

\end{document}